\documentclass[twocolumn]{wlscirep}
\usepackage[utf8]{inputenc}
\usepackage{graphicx}
\usepackage{amsmath}

\newcommand       \kg            {\,{\rm kg}}

\newcommand       \m        {\,{\rm m}}
\newcommand       \mm        {\,{\rm mm}}
\newcommand       \nm        {\,{\rm nm}}
\newcommand       \cm         {\,{\rm cm}}
\newcommand       \mum      {\,{\rm \mu m}}
\newcommand       \km         {\,{\rm km}}
\newcommand       \AU         {\,{\rm au}}
\newcommand       \au         {\,{\rm au}}
\newcommand       \rh         {{r_h}}
\newcommand       \s            {\,{\rm s}}
\newcommand       \GHz        {\,{\rm GHz}}

\newcommand       \microJy    {\,\mu{\rm Jy}}
\newcommand       \K            {\,{\rm K}}
\newcommand       \asec {^{\prime\prime}}
\newcommand     \gtsim  {\lower.5ex\hbox{$\buildrel > \over \sim$}}
\newcommand     \ltsim  {\lower.5ex\hbox{$\buildrel < \over \sim$}}
\newcommand     \simgt  {\lower.5ex\hbox{$\buildrel > \over \sim$}}
\newcommand     \simlt  {\lower.5ex\hbox{$\buildrel < \over \sim$}}
\newcommand       \simali       {{\sim\,}}
\newcommand       \Mdust       {M_{\rm dust}}
\newcommand       \rhoCD       {\rho_{\rm CD}}
\newcommand       \amin       {a_{\rm min}}
\newcommand       \amax       {a_{\rm max}}

\title{Pebbles and volatile evolution in the interstellar comet 2I/Borisov}

\author{Bin Yang$^{1,*}$, Aigen Li$^{2}$, Martin A. Cordiner$^{3,4}$, Chin-Shin Chang$^{5}$, Olivier R. Hainaut$^{1}$, Jonathan P. Williams$^{6}$, Karen J. Meech$^{6}$, Jacqueline V. Keane$^{6}$, and Eric Villard$^{5,}$}

\affil[1]{European Southern Observatory, Alonso de Córdova 3107, Vitacura, Santiago, Chile}
\affil[2]{Department of Physics and Astronomy, University of Missouri, Columbia, MO 65211, USA}
\affil[3]{Astrochemistry Laboratory, NASA Goddard Space Flight Center, 8800 Greenbelt Road, Greenbelt, MD 20771, USA}
\affil[4]{Department of Physics, Catholic University of America, Washington, DC 20064, USA}
\affil[5]{Joint ALMA Observatory, Alonso de Córdova, 3107, Vitacura, Santiago 763-0355, Chile}
\affil[6]{Institute for Astronomy, University of Hawaii, 2680 Woodlawn Drive, Honolulu, HI 96822, USA}
\affil[*]{byang@eso.org}

\begin{abstract}
 The interstellar traveler, 2I/Borisov, is the first clearly active extrasolar comet, ever detected in our Solar system. We obtained high-resolution interferometric observations of 2I/Borisov with the Atacama Large Millimeter/submillimeter Array (ALMA), and multi-color optical observations with the Very Large Telescope (VLT) to gain a comprehensive understanding of the dust properties of this comet. We found that the dust coma of 2I/Borisov consists of compact ``pebbles'' of  radii exceeding $\simali$1$\mm$, suggesting that the dust particles have experienced compaction through mutual impacts during the bouncing collision phase in the protoplanetary disk. We derived a dust mass loss rate of $\gtsim$200$\kg\s^{-1}$ and a dust-to-gas ratio $\gtsim$3. Our long term monitoring of 2I/Borisov with VLT indicates a steady dust mass loss with no significant dust fragmentation and/or sublimation occurring in the coma. We also detected emissions from carbon monoxide gas (CO) with ALMA and derived the gas production rate of $Q({\rm CO})=(3.3\pm0.8)\times10^{26}\s^{-1}$. We found that the CO/H$_2$O mixing ratio of 2I/Borisov changed drastically before and after perihelion, indicating the heterogeneity of the cometary nucleus, with components formed at different locations beyond the volatile snow-line with different chemical abundances. Our observations suggest that 2I/Borisov’s home system, much like our own system, experienced efficient radial mixing from the innermost parts of its protoplanetary disk to beyond the frost line of CO.

\end{abstract}
\begin{document}

\flushbottom
\maketitle
%
%
Planetary systems are born out of interstellar clouds of gas and dust grains, where dust plays an important role in radiative cooling of collapsing clouds as well as serving as seeds for condensation and accretion of the building blocks of planetary bodies. Dust particles are rich in information on their formation and evolution history, such as the transport and collisional processes in protoplanetary disks. However, it is nearly impossible to perform in depth investigations on the dust properties around other stars due to the large distances and faintness of these systems. 

Interstellar objects (ISOs) are planetesimals, the building blocks of planets, kicked out of their native planetary systems \cite{Charnoz2003}. Some of these interstellar wanderers eventually pass through our Solar system, providing us rare opportunities to characterize exo-planetesimals in unprecedented detail. The first ISO, 1I/`Oumuamua, was discovered in 2017 and exhibited a point-like appearance with no sign of cometary activity \cite{meech2017,ISSI2019}. In contrast, the second ISO, 2I/ Borisov, which was discovered in August 2019, unambiguously exhibited a coma and tail upon discovery \cite{Guzik2020}. The detection of typical cometary emissions such as CN (ref.\cite{Fitzsimmons2019}) and C$_2$ (ref.\cite{Opitom2019}) makes it the first obviously active extrasolar comet ever detected in our Solar system. In addition, this comet is rich in supervolatile CO \cite{Bodewits2020, Cordiner2020}, regardless of its small nucleus \cite{Jewitt2020,Hui2020}, which indicates that the nucleus of 2I/Borisov is likely to be pristine.

Solar system comets consist of ices and dust, where the major ice species are H$_2$O, CO, and CO$_2$ (ref.\cite{Cochran2015}) and the dust generally consists of silicates, oxides, and sulfides, as well as high-molecular weight refractory organics and amorphous carbon materials \cite{Fray2017,Levasseur2018}. When a comet enters the inner solar system, a coma of dust and gas and/or tails of dust and plasma begin to develop around the nucleus due to sublimation of surface ices in the heat of the Sun and the embedded dust particles are dragged out by the expanding gas \cite{Ahearn:2017}. In recent years, in-situ observations of comet 67P/Churyumov-Gerasimenko (hereafter 67P) by the ESA Rosetta spacecraft greatly enhanced our understanding of comets. The nucleus of 67P is thought to be a primordial rubble pile\cite{Davidsson2016} and its dust particles have an irregular, fluffy structure \cite{Levasseur2018}, with sizes varying widely from $\simali$1$ \mum$ to nearly 1\,m (ref.\cite{Fulle2017,guttler2019}). Rosetta measurements suggest that comets in our Solar system formed in a wide region beyond proto-Neptune and were scattered by giant planet migration to their present reservoirs\cite{Ahearn:2017}. 

\begin{figure*}[t]
\centering
\includegraphics[width=17cm]{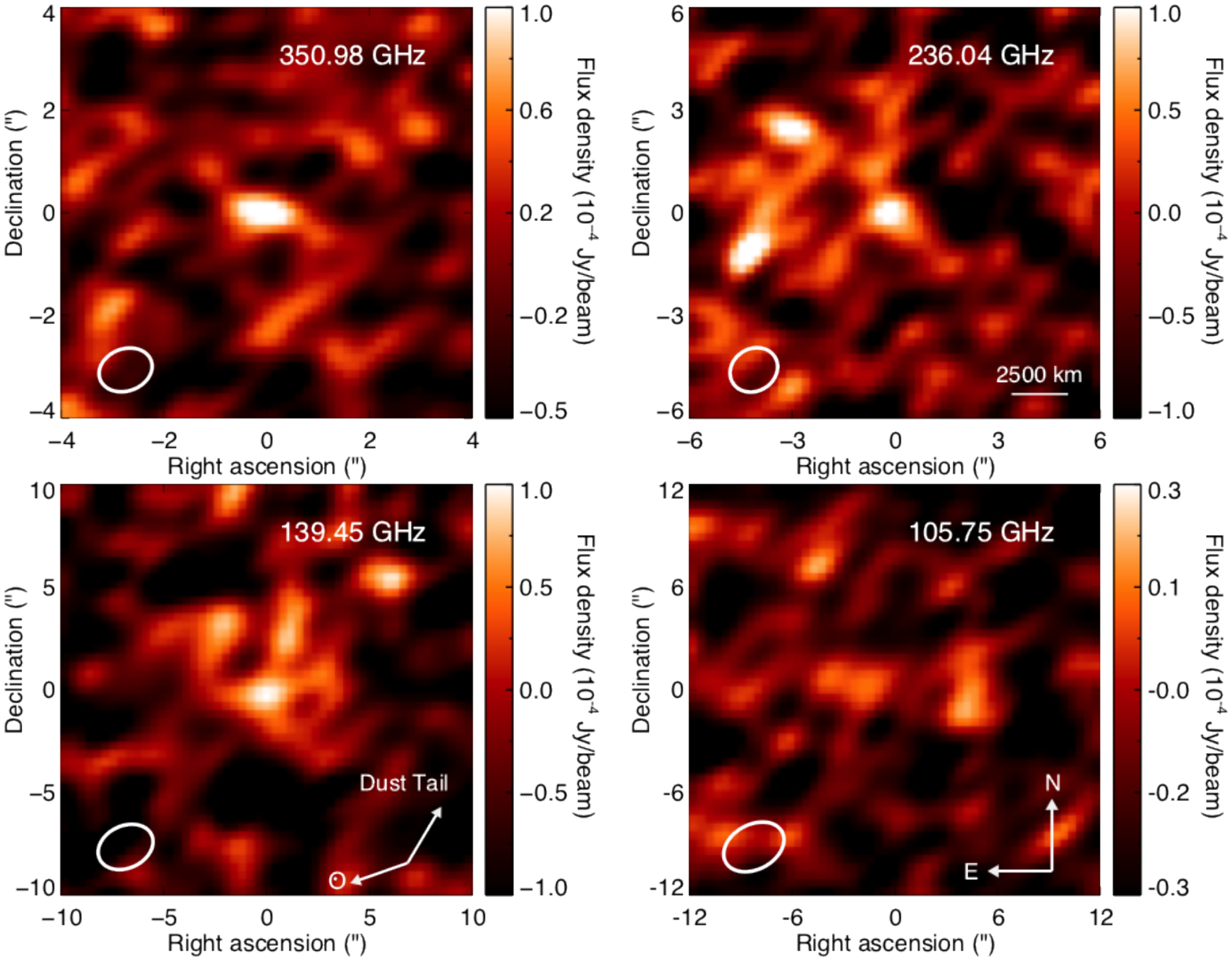}
\caption{ALMA dust thermal continuum maps of 2I/Borisov at 0.85, 1.27, 2.15 and 2.84$\mm$. The observations of the dust thermal continuum emission for the coma around the ISO 2I/Borisov were made at a heliocentric distance of $\rh\approx2.01\AU$ and a geocentric distance of $\Delta\approx2.03\AU$. The color-bar indicates the thermal emission intensity in Jansky. The projected position angles of the extended Sun-to-target vector and dust-tail orientation on 2019 December 03 are labeled on the lower right. The FWHM of the spatial resolution element for each band is labeled on the lower left.}
\label{fig:dust}
\end{figure*}

We obtained high angular resolution continuum observations of 2I/Borisov with ALMA in Bands 3 and 4 (2019 Dec. 03) and Bands 6 and 7 (2019 Dec. 02), respectively, which is less than a week before the comet reached perihelion at $\rh=2.007\AU$ on 2019 December 8. The ALMA dust continuum images of 2I/Borisov from 850$\mum$ to 2.84$\mm$ are shown in Fig.~\ref{fig:dust}. We detected the dust continuum emission at the center of the field of view (FOV) in three out of four bands, at 350$\GHz$ (Band 7), 236$\GHz$ (Band 6) and 139$\GHz$ (Band 4), respectively. In Band 4 and Band 6, there are some non-negligible structures near the center, which could be associated with the comet. However, due to the limited signal-to-noise ratio of these structures, we excluded them from further analysis and only measured the flux at the center of the image within one ALMA beam, where the strongest signal was detected. 

We modeled the measured thermal flux of 2I/Borisov, using aggregates of amorphous silicates and amorphous carbon grains (see Methods) \cite{Li1998}. We found that very large, ``pebble''-like {\it compact} particles of radii $a\simgt1\mm$ are required to reproduce the ALMA photometry. Here we follow the official terminology of the {\it United States Geological Survey} (USGS) for which ``pebbles'' refer to mm- to cm-sized particles \cite{Wentworth1922}, which are smaller than those cm- to meter-sized pebbles typically considered in the ``pebble accretion'' growth of planetesimals \cite{Johansen2017}.

\begin{figure*}[!t]
\centering
\includegraphics[width=1.6\columnwidth]{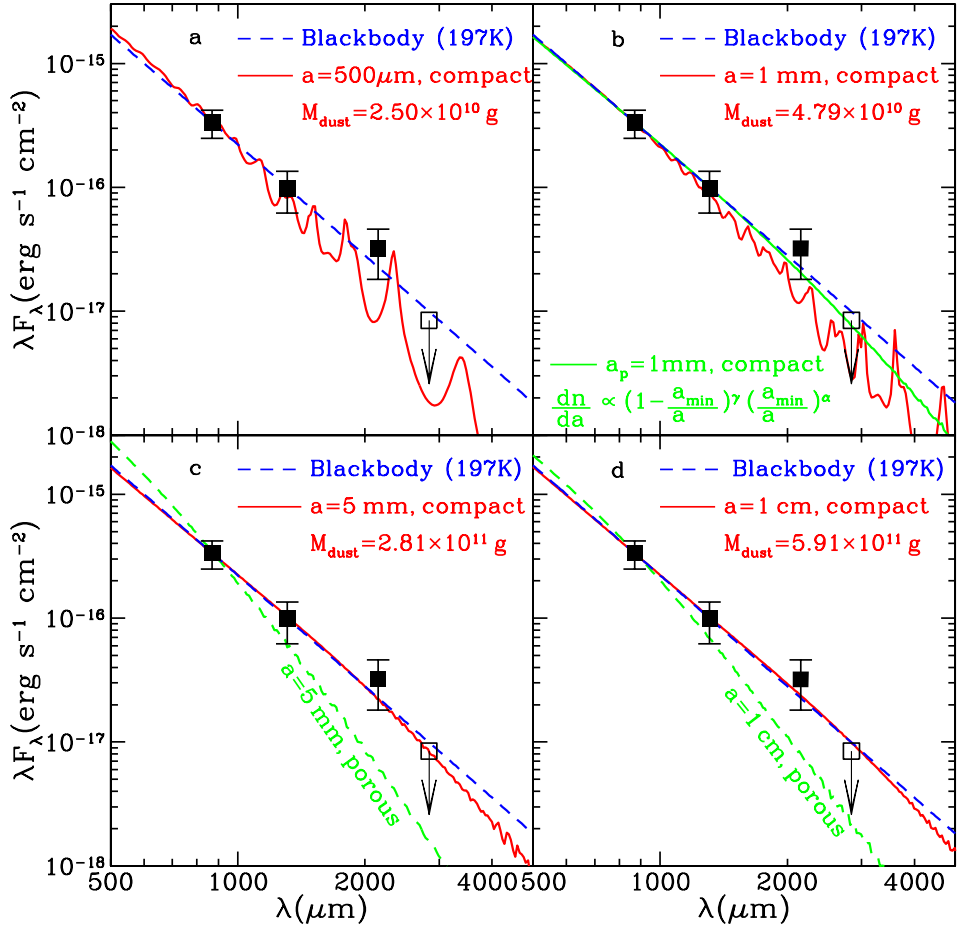}
\vspace{-35mm}
\caption{Dust thermal emission models and ALMA photometry of 2I/Borisov. The ALMA photometric data of comet 2I/Borisov are shown as black squares for Band 7 (350$\GHz$, 0.85$\mm$), Band 6 (236$\GHz$, 1.27$\mm$), and Band 4 (139$\GHz$, 2.15$\mm$). The comet is not detected at Band 3 (106$\GHz$, 2.84$\mm$) and the flux measurement at the 1$\sigma$ upper limit is shown as an open square. Arrow indicates 1$\sigma$ upper limits, and error bars indicate 1$\sigma$ uncertainties. Blue dashed lines shows the blackbody fit for a temperature of $T=197\K$, the equilibrium temperature expected for pebble-like, mm-sized (and larger) particles at $\rh=2.01\AU$. Red solid lines show the model SED calculated from compact, spherical particles of amorphous silicates and amorphous carbon of {\it single} radii of 500$\mum$ (a), 1$\mm$ (b), 5$\mm$ (c) and 1$\cm$ (d). For particles of sizes $a$\,=\,500$\mum$ (a) and 1$\mm$ (b), the model SEDs exhibit pronounced resonant fine structures due to interference effects \cite{Bohren:1983}, 
which will be smoothed out if dust size distributions are considered. Indeed, the model SED arising from compact, spherical particles of a Divine-type size distribution \cite{Divine1986} of which the size peaks at $a_p=1\mm$ and the area-weighted size peaks at $\simali$2.3$\mm$ (green solid line in (b)) is smooth and featureless (see Methods). For particles of sizes $a$\,=\,5$\mm$ (c) and 1$\cm$ (d), the model SEDs do not show any appreciable resonant structures in the ALMA wavebands because these pebbles are approaching the geometrical optics regime of $2\pi a/\lambda$\,$\gg$\,1 (see ref.\cite{Bohren:1983}). Also shown are the SEDs calculated from porous particles of porosity $P=0.80$ (green dashed lines) and of radii 5$\mm$ (c) and 1$\cm$ (d).}
\label{fig:sedmod}
\end{figure*}

As shown in Fig.~\ref{fig:sedmod}, smaller particles with $a<1\mm$ or larger {\it porous} particles with $a>1\mm$ would emit too little at $\lambda\gtsim1300 \mum$ to account for the ALMA photometry. As a matter of fact, the ALMA photometry closely resembles blackbody emission. In comparison, most Solar system comets exhibit a steeper spectral energy distribution (SED) in the submm and mm wavelength range (e.g., see ref. \cite{jewitt1999}). To emit like a blackbody, these dust particles have to be much larger than the wavelength (i.e., $a\gg\lambda/2\pi$; see ref.\cite{Bohren:1983}) which, at the ALMA wavebands, is indicative of mm-sized (or larger) pebbles. These pebbles are in the geometrical optics regime so that their absorption cross sections are essentially constant from the ultraviolet all the way to the submm and mm wavelengths \cite{Bohren:1983}. The dust particles attain an equilibrium temperature of $T\approx 197\K$ at $\rh=2.01\AU$ and their emission spectra are blackbody-like (see Methods). We have also fitted the SED in terms of a Divine-type dust size distribution \cite{Divine1986} which also suggests the predominant presence of pebble-like dust particles of $a\simgt1\mm$ (see Fig.~\ref{fig:sedmod}).
Other ground-based optical observations yielded an effective dust radius from 100 $\mum$ up to 3$\mm$ for 2I/Borisov (Oct. 2018 -- May 2019\cite{Ye2020}, Sept. -- Oct. 2019\cite{jewitt2019}, Nov. -- Dec. 2019\cite{Cremonese2020}). Hubble Space Telescope (HST) observations of this ISO near its perihelion reveal a slight asymmetry in the dust coma that requires the presence of mm-sized particles \cite{Kim2020}. 


Assuming an effective dust size of $\simali$5$\mm$, we obtained a total dust mass loss of $\approx2.81\times10^{8}\kg$ within the ALMA beam (see Fig.~\ref{fig:sedmod}). Larger pebbles (e.g., $a\simgt1\cm$) emit less effectively at the ALMA bands and cannot be well constrained, therefore, our dust mass loss estimate is considered as a lower limit (see discussion in Methods). Adopting a dust velocity of 3$\m\s^{-1}$ for mm-sized particles around perihelion (ref.\cite{Cremonese2020}) and an effective beam size of 2.8$^{\prime\prime}$ (projected distance of $4.12\times10^3\km$), we obtained a beam crossing time $\tau$\,$\simali 1.36\times10^{6}\s$, in turn, a dust mass loss rate $Q({\rm dust})\sim200\kg\s^{-1}$. Similar results were obtained if we considered the Divine-type dust size distribution which peaks at $a_p=1\mm$ and requires a total dust mass loss of $\approx2.57\times10^{8}\kg$ (see Methods). 

In complement to the ALMA observations, we also monitored the dust activity of 2I/Borisov with the FOcal Reducer and low dispersion Spectrograph 2 (FORS2) on VLT from November 2019 until March 2020. Using the FORS observations, we measured and calculated the $Af\rho$ quantities and derived order-of-magnitude dust mass loss rates (see Methods), which are listed in Supplementary Table~2. We note that the dust mass loss rate obtained using ALMA is at least four times higher than the rates derived from optical observations (our FORS observations and ref.\cite{Kim2020,Cremonese2020}).
While the $Af\rho$-based $Q({\rm dust})$ rates derived from the FORS observations are just order-of-magnitude estimates, it is actually not surprising that the ALMA-based $Q({\rm dust})$ rate considerably exceeds the $Af\rho$-based rates. It is well recognized that the bulk of the particulate coma mass is typically carried by large dust particles 
and optical observations preferentially sample the smaller particles since they are typically more abundant and contribute to a larger fraction of the optical cross-sections than do the mass-dominant larger particles \cite{Li1998}.

Our FORS monitoring observations show that the dust mass loss rates only exhibit small variations with respect to heliocentric distances during the monitoring period. Other observations \cite{Ye2020,Cremonese2020} and dynamical dust modelling also indicate that the dust production rate has been quite stable for weeks, possibly months before the observations. We therefore adopted a constant dust production rate of 200$\kg\s^{-1}$ and obtained the total mass loss of $2.0\times 10^{9}\kg$ during the period between the discovery and the arrival at the perihelion. Adopting a typical bulk density of 500$\kg\m^{-3}$ for the nucleus \cite{Patzold2019} and assuming uniform erosion of the nucleus surface (with $r_n=0.4\km$ for the nucleus radius; see ref.\cite{Hui2020}), we find that a layer of at least 2$\m$ was eroded from 2I/Borisov's surface since it entered the inner Solar system in late August 2019 . 

\begin{figure}
    \includegraphics[width=1.0\columnwidth]{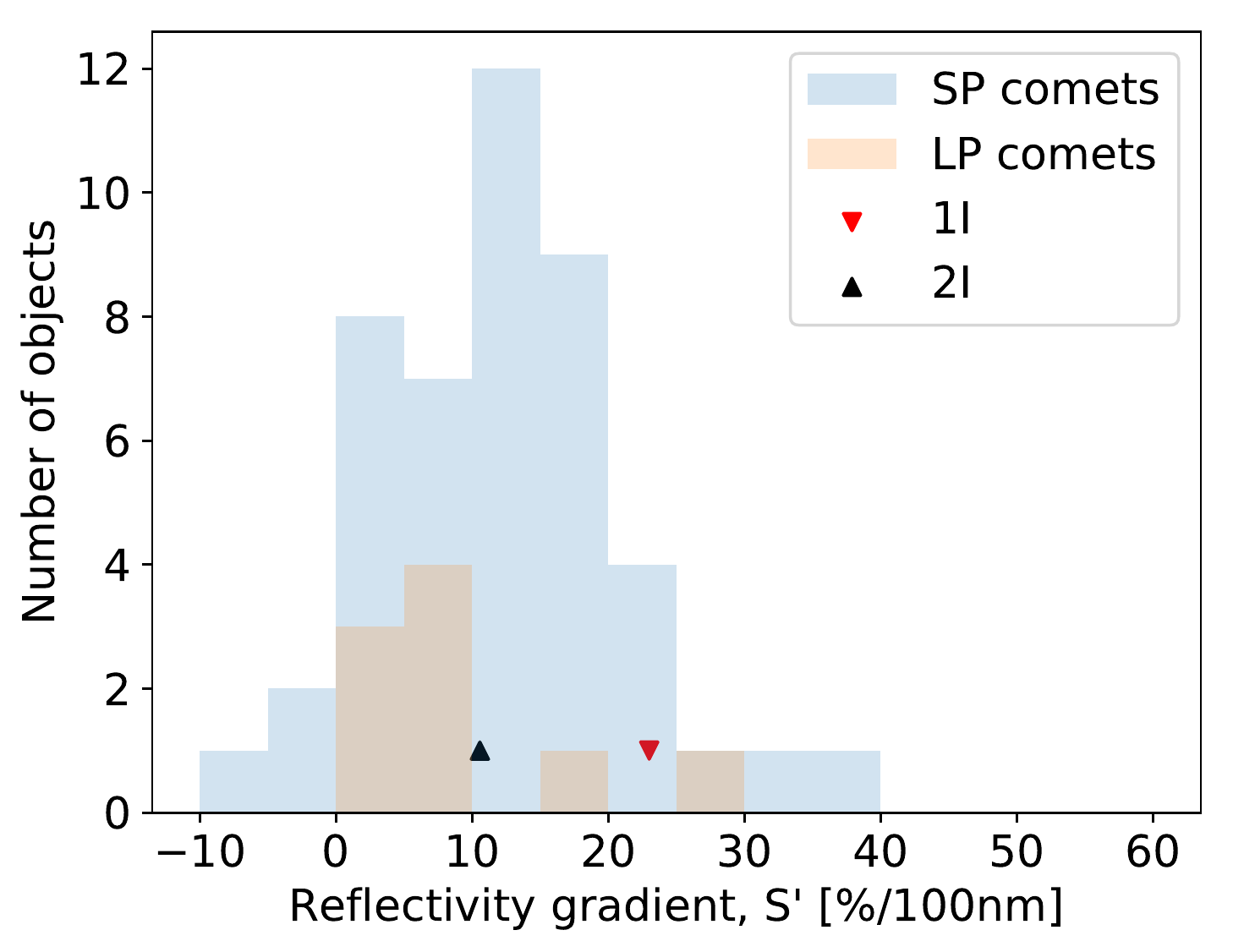}
    \caption{Optical reflectivity gradient of 2I/Borisov in comparison to 1I/`Oumuamua \cite{meech2017} and solar system comets. \cite{hainaut12MBOSS} The reflectivity gradient histograms are for short- and long-period comets from the solar system \cite{hainaut12MBOSS}.}
    \label{fig:gradientHist}
\end{figure}

 Our spectroscopic and multi-band photometric observations with FORS show that the dust coma of 2I/Borisov appears slightly redder than the Sun in the optical with an average reflectivity $S^{\prime}=(9\pm5)\%/100\nm$, consistent with other studies \cite{Hui2020,Bolin2020a}. In comparison, the surface of 1I/'Oumuamua had a much redder slope of $(23\pm3)\%/100\nm$ (see ref.\cite{meech2017}), as illustrated in Fig.~\ref{fig:gradientHist}. We found that the dust color and the dust coma profile showed no sign of evolution with time. For cometary dust, the optical color depends mainly on the particle size distribution, rather than on the dust composition \cite{Kolokolova1997}. The absence of temporal and spatial color variations indicates that no significant dust fragmentation and/or sublimation from grains occurred in the coma of 2I/Borisov, which is consistent with the lack of icy particles in the coma\cite{Yang2020}.  
 
\begin{figure*}[h!]
    \centering
    \includegraphics[width=15cm]{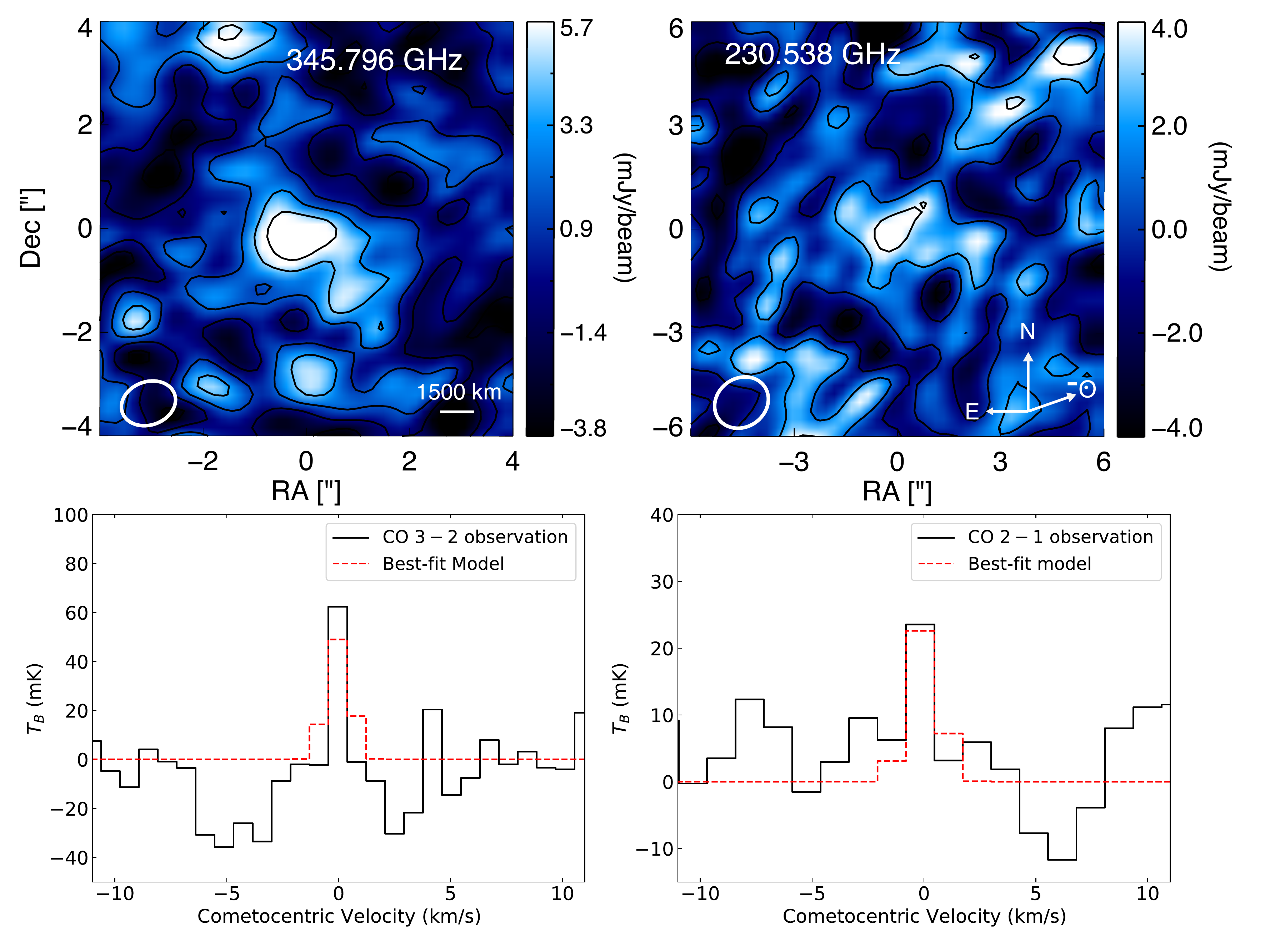}
    \caption{CO flux maps and spectra of 2I/Borisov obtained with ALMA. 
    The CO emissions were detected at the center of each image in ALMA Band 7 (upper left) and Band 6 (upper right), respectively. The ALMA interferometric spectra of CO (J\,=\,3--2) (lower left) and CO (J\,=\,2--1) (lower right) towards 2I/Borisov were extracted at the respective emission peaks of each species. The best-fitting radiative transfer models are overlaid (see Methods).
    }
    \label{fig:gas_co}
\end{figure*}

Besides the dust continuum, we also detected gas emission from the carbon monoxide (CO) $J=3-2$ and $J=2-1$ transitions in ALMA Bands 6 and 7, respectively (see Fig.~\ref{fig:gas_co}). The emission was confined to a single spectral channel in each case, centered on the radial velocity of the nucleus. The two CO lines were modeled simultaneously, allowing the temperature $T$ and production rate $Q(\rm CO)$ to vary as free parameters until the best fit to the observed spectra was obtained. The best-fitting model gives $Q({\rm CO})=(3.3\pm0.8)\times10^{26}\s^{-1}$ and $T\approx23\K$. However, the temperature is not well constrained by the observations, with a ($1\sigma$) error range of $\simali$13--181$\K$. Using a water production rate $Q({\rm H_2O})=(10.7\pm1.2)\times10^{26}\s^{-1}$ on 01 December 2019 (see ref.\cite{xing2020}), we derive a CO/H$_2$O mixing ratio of $(31\pm8)\%$.

 Around perihelion, the water production rate is $\simali$30$\kg\s^{-1}$ (see ref.\cite{xing2020}) and the CO production rate is $\simali$15--20$\kg\s^{-1}$ (this study; see also ref.\cite{Cordiner2020}). If we assume the CO$_2$ production rate is comparable to that of CO, similar to the case of 67P (ref.\cite{lauter2019}), then the total gas production rate is $\simali$60--70$\kg\s^{-1}$. Our ALMA observations indicate that the dust mass loss rate is $\gtsim 200 \kg\s^{-1}$, in turn, the dust-to-gas ratio of 2I/Borisov is $\chi_{\rm 2I} \gtsim 3$. In comparison, comet C/1995 O1 (Hale-Bopp) has $\chi_{\rm HP}>5.0$ (see ref.\cite{jewitt1999}) and 67P has $\chi_{\rm 67P}$\,$\simali$0.6--6 (see refs.\cite{Rotundi2015, Fulle2017,Choukroun2020}), while most solar system comets exhibit a dust-to-gas ratio around 1 or less\cite{Singh1992,Lamy2009}. Theoretical studies have shown that for comets formed through the gravitational collapse of pebble clouds, their dust-to-gas ratios would be in the range of $3 \lesssim \chi \lesssim 9$ (ref.\cite{Lorek2016}). The relatively high dust-to-gas ratio of 2I/Borisov is consistent with the scenario of its formation in a collapsing pebble cloud.


While Solar system cometary dust particles are largely comprised of irregular, fluffy aggregates \cite{Levasseur2018,Mannel2019}, the predominance of compact particles in the coma of 2I/Borisov suggests that the dust particles of this ISO have experienced mutual impacts during the bouncing collision phase in the protoplanetary disk \cite{Weidenschilling1977, Zsom2010}, which led to the reduction of the empty space inside the fluffy particles and ultimately restructuring fractal particles into pebbles \cite{Zsom2010, Fulle2017}. The high collision rate suggests that the large dust particles in 2I/Borisov must have formed in the inner protoplanetary disk of its parental star. The compaction of porous particles could also result from the packing forces produced by an anisotropic sublimation of the surface layers of loosely conglomerated fluffy aggregates\cite{Mukai1983}.

Comparing the pre-perihelion CO production rate obtained in this study to other CO production rates obtained post-perihelion\cite{Bodewits2020, Cordiner2020}, it is apparent that the CO sublimation rate of 2I/Borisov varied significantly around perihelion, and is decoupled from the water production. As discussed in refs.\cite{xing2020, Bodewits2020}, the release of water peaked when 2I/Borisov was closest to the Sun. After perihelion, the water production dropped sharply by a factor of 5 in 20 days \cite{xing2020}. In contrast, the CO production rate increased steadily as the comet approached the Sun, and reached the peak about a month after perihelion. The abundance of CO with respect to H$_2$O increased from about 30\% (pre-perihelion) to nearly 160\% (post-perihelion), making 2I/Borisov the most CO-rich comet observed within 3$\AU$ of the Sun \cite{Bodewits2020, Cordiner2020}.

The decoupled CO and H$_2$O sublimation pattern as well as the large variation in CO to H$_2$O ratio has been observed previously in solar system comets. For instance, the CO rich comet C/2009 P1 (Garradd) showed H$_2$O production peaking $\simali$50 days before perihelion while the CO production rate increased monotonically until 100 days post-perihelion \cite{Feaga2014}. The asymmetric activity of C/2009 P1 is attributed either to a seasonal effect, where the sublimation is coming from two separate active regions that have different compositions, or to the exposure of different sub-layers of the nucleus, with differing CO/H$_2$O ratios \cite{Feaga2014}. As discussed earlier, the surface layer of the nucleus of 2I/Borisov had eroded at least two meters when it reached the perihelion. The thickness of eroded materials is comparable to the depth, in the range of 1--10\,m, of the devolatilized mantle that is heavily processed by cosmic-ray irradiation or supernova heating \cite{Cooper2003,Garrod2019}. The post-perihelion increase in CO activity could be due to the exposure of the volatile-rich deeper layers as a result of surface erosion. However, unveiling sub-surface fresh materials is not compatible with the steep decline in the post-perihelion water production rate. In addition, our FORS spectroscopic observations found that the production rates of other species, i.e., CN, C$_2$, NH$_2$, also followed the H$_2$O production rate and dropped steadily after perihelion\cite{Jehin2020}. 

Rosetta observations of 67P revealed that the peak of water flux was confined to the subsolar latitude either on the northern hemisphere or the southern hemisphere, whereas, the hypervolatile CO$_2$ and CO gases were not specifically originating from the sunlit regions \cite{lauter2019}. Two studies based on the HST observations of 2I/Borisov suggest that the sub-solar latitude of the nucleus evolved either from $-35^\circ$ to 0$^\circ$ or from $-10^\circ$ to 70$^\circ$ between August 2019 and January 2020 \cite{Kim2020, Bolin2020b}. The drastic drop in the water production rate of 2I/Borisov may have been due to the possibility that water sublimation was mainly coming from a confined region on the surface of the nucleus. Post perihelion, this region rotated away from the sun, leading to decreasing water sublimation rates. Consistently, post-perihelion ALMA observations revealed that the source of HCN (possible parent molecule of CN) outgassing was confined to a narrow region in the plane of the sky, whereas CO outgassing was more isotropic \cite{Cordiner2020}. The evolution of CO/H$_2$O ratio is unlikely due to exposing fresh subsurface materials and, therefore, does not represent the pristine abundance ratio. Instead, the large variation in the CO/H$_2$O ratio reflects a heterogeneity of the cometary nucleus, most likely with components formed at different locations beyond the volatile snowline, with differing chemical abundances. 
 
Our ALMA and VLT observations indicate that 2I/Borisov's home planetary system, much like our own Solar system, had experienced efficient radial mixing from the innermost parts of its protoplanetary disk to beyond the frost line of CO. Among a number of probable mechanisms that have been proposed for the origin of ISOs \cite{ISSI2019}, gravitational interactions between planetesimals in the protoplanetary disk and growing giant planets is favored, as it can explain both the ejection of ISOs from their home systems as well as account for the strong radial transport of materials in the disk \cite{walsh2011}. While the most common planets in other exoplanetary systems seem to be super-Earths and mini-Neptunes \cite{Batalha2014}, our study suggests the presence of giant planets in the home system of 2I/Borisov. 

\section*{Methods}
2I/Borisov was observed with ALMA during Cycle 7, under the Director's Discretionary Time program 2019.A.00002.S. The observations were carried out 5 and 6 days before perihelion for Bands 3 and 4 (UT 2019-12-03) and Bands 6 and 7 (UT 2019-12-02), respectively. Our observations were performed using the C43-2 nominal configuration, spanning baselines 15--161$\m$, which provided an angular resolution of $1^{\prime\prime}$ ($4^{\prime\prime}$) at 350$\GHz$ (109$\GHz$). The flux scale was calibrated with respect to the quasar J1130-1449. 

For the continuum observations, we used the largest bandwidth of 7.5$\GHz$, by combining the four 1.875$\GHz$ base-bands in dual-polarization mode. The correlator was configured to observe rotational transitions of CO $J$\,=\,2--1 (230.538$\GHz$) in Band 6 and CO $J$\,=\,3--2 (345.796$\GHz$) in Band 7, with a spectral resolution of 1.128~MHz and a channel spacing of 977\,kHz. The data reduction was performed using CASA version 5.6.1-8 and ALMA Pipeline version r42866, and the imaging was performed using the CASA $\tt tclean$ task with natural weighting. Since we were tracking on the comet at a rate (0.03$^{\prime\prime}$/s) throughout the observations, any background sources are expected to be smeared in the final images. As shown in Fig.~\ref{fig:dust}, there are some structures near the center of the image, close to the expected location of the comet. Given the high sensitivity and spatial resolution of ALMA at the time of the observation, it is possible that some dust coma structure can be resolved. However, due to the low signal-to-noise ratio of these features, we can not confirm the extended features are real features. Therefore, we only measure the flux at the very center of each image where the strongest signal was detected.   

Due to the wide wavelength coverage, the image resolution/beam size is significantly different ($\theta_7 = 1.08\asec\times0.84\asec$, $\theta_6 = 1.53\asec\times1.31\asec$, $\theta_4 = 2.84\asec\times2.09\asec$, $\theta_3 =  4.05\asec\times2.83\asec$), as shown in Fig.~\ref{fig:dust}. To make a consistent measurement of the continuum flux over the four bands, we convolved all the continuum images with a fixed resolution element of $3.4^{\prime\prime}\times 2.3^{\prime\prime}$. We integrated total flux within one FWHM of the resolution element in the re-convolved images. To estimate the noise level, we measured the root mean square (RMS) within one beam at multiple locations that are close to the center for each band, using the CASA viewer $statistic$ task. We then calculated the mean of the measured RMS values as the final noise estimate for each band. The measured total flux as well as the noise in each band is listed in Table~\ref{tab:flux}.
\begin{table}[h]
\begin{tabular}{ccccccc}
    \hline\hline
    Band & Freq. & Itime & $\Theta$ & Flux & RMS & $\sigma$ \\
    &
     (GHz) & (min) & ($\asec$) & ($\microJy$) & ($\microJy$) & \\ 
    \hline
 7 & 350 & 41 & 3.4$\times$2.3  & 94 & 25 & 3.8\\
 6 & 236 & 34 & 3.4$\times$2.3  & 44 & 15 & 2.9 \\
 4 & 139 & 55 & 3.4$\times$2.3  & 21 &  7 & 3.0 \\
 3 & 106 & 72 & 3.4$\times$2.3  &  8 &  6 & 1.3 \\
\hline
\end{tabular}
  \caption[Photometry of ALMA data]{Photometric measurements of the thermal emission of the dust coma over four ALMA bands. $Itime$ is a total on-source observing time in minutes and $\Theta$ is a spatial resolution element (elliptical beam FWHM). The total flux is summed up within one ALMA beam for each band. The contributions of the nucleus to the fluxes at Bands 7, 6, 4 and 3 are at most $\simali$3.73, 1.65, 0.64 and 0.36 $\microJy$, respectively, for a nucleus upper-limit radius of $\simali$0.4$\km$ (ref.\cite{Hui2020}) and a Bond albedo of 0.04 which corresponds to an effective temperature of $\simali$195$\K$ at $\rh=2.01\au$.}
  \label{tab:flux}
\end{table}

We model the dust thermal emission of 2I/Borisov in terms of cometary dust paricles---porous aggregates 
of amorphous silicates and amorphous carbon grains
\cite{Li1998}---which are commonly used for 
modeling the dust-scattered sunlight 
and dust thermal emission of comets 
\cite{Kolokolova2004, Kimura2016}.
The dust is characterized by 
(i) size $a$ -- the radius of the sphere encompassing the entire aggregate 
(we assume that all grains are spherical in shape),
(ii) porosity $P$ -- the fractional volume of vaccum, and
(iii) mixing mass ratio $m_{\rm carb}/m_{\rm sil}$
for the amorphous silicate and amorphous carbon constituent grains. 
We use Mie theory combined with the Bruggman effective medium theory \cite{Bohren:1983} to calculate $C_{\rm abs}(a,\lambda)$, the absorption cross sections of spherical porous aggregates of radii $a$ at wavelength $\lambda$. We take the dielectric functions of Draine \& Lee\cite{Draine:1984zt} for amorphous silicate dust and of Li \& Greenberg \cite{Li1997} for amorphous carbon dust. We adopt $m_{\rm carb}/m_{\rm sil}$\,=\,0.5, as implied from the mass spectrometry of the dust particles from 67P measured with the {\it Cometary Secondary Ion Mass Analyzer} (COSIMA) aboard Rosetta \cite{Bardyn2017}. The chemical composition of the dust grains from Halley as inferred from the impact-ionization time-of-flight mass spectrometry \cite{Jessberger1988} suggested $m_{\rm carb}/m_{\rm sil}$\,$\approx$\,0.8. We note that, since the observed SED in the ALMA bands closely resembles blackbody emission (see Fig.~\ref{fig:sedmod}), the exact dust composition is less relevant. Indeed, our calculations show that both pure silicate particles and pure amorphous carbon particles can reproduce the ALMA observations, provided that they are mm-sized (or larger) pebbles.

At a heliocentric distance $\rh$, a spherical dust particle of radius $a$ attains an equilibrium temperature of $T$ through balancing the absorption with the emission:
\begin{equation}\label{eq:Td}
\begin{split}
\int^\infty_0 C_{\rm abs}(a,\lambda)
\left(\frac{R_{\odot}}{2\rh}\right)^2
            F^{\odot}_\lambda d\lambda \\
            = \int^\infty_0 C_{\rm abs}(a,\lambda)
4\pi B_\lambda\left(T[a,\rh]\right)d\lambda~,
\end{split}
\end{equation}
where $R_{\odot}\approx4.65\times10^{-3}\AU$ is the solar radius, $F^{\odot}_\lambda$ is the flux per unit wavelength (${\rm erg}\s^{-1}\cm^{-3}$) at the top layer of the solar atmosphere, and $B_{\lambda}(T)$ is the Planck function at temperature $T$ and wavelength $\lambda$. We note that in eq.\,\ref{eq:Td}, the factor 1/2 arises from the ``dilution'' factor
%
\begin{equation}\label{eq:omega}
\omega = \frac{1}{2}
\left\{1-\sqrt{1-\left(R_\odot/\rh\right)^2}\right\}~~,
\end{equation}
which measures the factor by which the energy density in the radiation field is reduced as the source
of radiation moves to a large distance. At the ``surface'' of the Sun (i.e., $\rh=R_\odot$), it is obvious that $\omega$\,=\,1/2 (actually a little less because of limb-darkening). At $\rh\gg R_\odot$,
$\omega$\,$\approx$\,$\left(R_\odot/2\rh\right)^2$. Assuming the coma consists of dust particles of a single size $a$ and a total mass of $\Mdust$, at a comet-Earth distance of $\Delta$, we calculate the dust flux density from 
\begin{equation}\label{eq:jnu}
F_{\lambda}(a,\rh) = \frac{4\pi B_{\lambda}(T[a,\rh])\,C_{\rm abs}(a,\lambda)}{4\pi \Delta^2} \times \frac{\Mdust}{\left(4\pi/3\right)\,a^3\rhoCD}~~~,
\end{equation}
where $\rhoCD$ is the mass density of cometary dust (for compact dust, $\rhoCD\approx2,650 \kg\m^{-3}$; for porous dust of porosity $P=0.80$, $\rhoCD\approx530\kg\m^{-3}$; see ref.\cite{Li2003}). We have also modeled the SED in terms of the Divine-type cometary size distribution \cite{Divine1986} $dn/da\propto \left(1-\amin/a\right)^{\gamma}\times\left(\amin/a\right)^{\alpha}$ in the size range of $\amin < a <\amax$. For such a function, the size distribution peaks at $a_p=\amin\times\left(\gamma+\alpha\right)/\alpha$ and the area-weighted size distribution is skewed towards larger particles and peaks at $a_p^{\prime}=\amin\times\left(\gamma+\alpha-2\right)/\left(\alpha-2\right)$. As far as the dust thermal emission is concerned, the area-weighted size peak $a_p^{\prime}$ is more meaningful than $a_p$ since $F_\lambda\propto C_{\rm abs}(a,\lambda)$ and $C_{\rm abs}(a,\lambda)\approx \pi a^2$ for dust particles in the geometrical optics limit.

By varying the lower and upper cutoff sizes ($\amin$, $\amax$) as well as the power exponents ($\alpha$, $\gamma$), close fits to the observed SED are achieved if the dust particles are predominantly mm-sized (or larger). As illustrated in Fig.~\ref{fig:sedmod}b, with $\alpha=3.5$ (see ref.\cite{Lasue2009}), $\amin=1\mum$ and $\amax=1\cm$, compact, spherical dust particles with a Divine-type size distribution peaking at $a_p=1\mm$ and a total dust mass of $M_{\rm dust}\approx2.57\times10^{8}\kg$ could reproduce the ALMA photometry fairly well. In contrast, dust particles with the same size distribution but peaking at $a_p=100\mum$ emit too little at $\lambda\gtsim1300\mum$. Compared with that calculated from dust particles of a single size $a=1\mm$ (also see Fig.~\ref{fig:sedmod}b), the model emission spectrum from those with the $a_p=1\mm$ Divine-type size distribution is smooth and does not show the resonant structures. These resonant structures seen in the emission spectra of single-sized particles arise from the interference between the incident and forward-scattered sunlight
which results in broad peaks in their absorption cross sections (see ref.\cite{Bohren:1983}). The positions of these interference-caused maxima and minima in the absorption cross sections are size dependent and therefore a size distribution would smear out these resonant structures. Also, the model SED calculated from dust particles with the $a_p=1\mm$ Divine-type size distribution agrees better with the ALMA photometry since the area-weighted size distribution is skewed to larger particles and peaks at $a_p^{\prime}\approx2.3\mm$.

Eq.\ref{eq:jnu} clearly shows that, for the model SED to resemble the blackbody-like SED observed by ALMA, the dust absorption cross sections $C_{\rm abs}(a,\lambda)$ have to be independent of wavelength in the ALMA bands, i.e., the dust particles have to be in the geometrical optics regime in the ALMA bands (and by implication, in shorter wavelengths as well) with $2\pi a/\lambda$\,$\gg$\,1 and $2\pi a\,|m-1|/\lambda$\,$\gg$\,1, where $m(\lambda)$ is the dust complex index of refraction \cite{Bohren:1983}. Under the geometrical optics condition, $C_{\rm abs}(a,\lambda)\approx\pi a^2$ and from eq.\ref{eq:Td} we derive $T(a,\rh)\approx 279\left(\rh/{\rm au}\right)^{-0.5}\K$, which corresponds to $T\approx197\K$ at $\rh=2.01\au$. The required dust mass linearly increases with the dust size: 
\begin{equation}\label{eq:dustmass}
\Mdust = 
\frac{\Delta^2\,F_\lambda^{\rm obs}}
     {B_\lambda(T[a,\rh])}
     \times 
     \left(\frac{4a\,\rhoCD}{3}\right)~,
\end{equation}
where $F_\lambda^{\rm obs}$ is the flux density observed by ALMA. We note that, for porous dust, the effective index of refraction $m(\lambda)$ is reduced and the geometrical optics condition of $2\pi a\,|m-1|/\lambda$\,$\gg$\,1 may not be met even if $2\pi a/\lambda$\,$\gg$\,1. This is why mm-sized, porous pebbles do not emit like blackbodies and fail to reproduce the observed SED (see Fig.~\ref{fig:sedmod}). The fits provided by smaller porous particles are even worse since they are even further away from the geometrical optics regime. 

We have so far assumed the dust particles to be spherical. However, cometary dust particles often have an irregular shape and fluffy structure \cite{Levasseur2018}. Unfortunately, there is no accurate solution to the absorption and scattering of light by mm-sized, irregular particles. The discrete dipole approximation, a powerful method for modeling the interaction between light and irregular particles, is limited to dust of sizes not much larger than the incident wavelength \cite{Draine1994}. Nevertheless, the assumption of spherical shapes (together with the Bruggeman or the Maxwell-Garnett effective medium theories for inhomogeneous grains) is sufficient in modeling the featureless mm and submm thermal continuum emission. 

We also note that Rosetta observed that a certain amount of large, decimeter-sized particles emitted from the nucleus eventually fall back onto the surface of 67P (ref.\cite{Fulle2017}), although the fraction of fall back is currently not well constrained. Given the smaller and lighter nucleus of 2I/Borisov, the cut-off  dust size should be larger than decimeters. As discussed earlier, dm sized particles are not sampled by ALMA observations. Therefore, we did not consider the fall-back effect when estimating the total dust mass loss. Our dust mass loss estimate as well as the dust-to-gas ratio are both considered as a lower limit.

 The integrated CO line flux is $\int{S_{\nu}}dv=4.8\pm1.2$\,mJy\,km\,s$^{-1}$\,beam$^{-1}$ for CO $J=3-2$ (which corresponds to a $4\sigma$ detection significance), and $\int{S_{\nu}}dv=2.8\pm1.0$\,mJy\,km\,s$^{-1}$\,beam$^{-1}$ for $J=2-1$ ($2.8\sigma$ significance). The CO observations were analyzed using a spherically symmetric coma radiative transfer model\cite{Cordiner2020}, expanding with a constant velocity, and subject to photolysis by solar radiation. Our data were of insufficient spectral resolution to determine the gas radial velocity based on the observed emission line profiles, so we adopted an outflow velocity of 0.47$\km\s^{-1}$, derived from spectrally resolved CO observations by Cordiner et al.\cite{Cordiner2020} on 15--16 December 2019. Spatial filtering of the model coma images by the interferometer was accounted for using the CASA {\tt simobserve} task, with the same hour angle, observation duration and array configuration as the science observations. Spectra of the two CO lines (extracted from their respective emission peak positions) were modelled simultaneously, allowing the temperature ($T$) and production rate ($Q$) to vary as free parameters until the best fit to the observed spectra was obtained (as determined using the reduced chi-squared statistic). Local thermodynamic equilibrium (LTE) was found to be applicable due to the relatively small ALMA beam size combined with the long lifetime of CO with respect to radiative transitions \cite{Cordiner2020}.

2I/Borisov was observed from Paranal using the FOcal Reducer and low dispersion Spectrograph 2 (FORS2) on ESO's Very Large Telescope (VLT), in service mode, from November 2019 until March 2020. The observation strategy depended on the epoch, serving various purposes: high-precision astrometry (taking a series of short exposures in the $R$ filter), deep imaging for studying the coma (stacking a long series of $R$ images), and for studying the colour of the object (alternating short images through the $b$, $v$, $R$, $I$, and $z$ filters). The journal of the observations is listed in Supplementary Table~1. The data were processed in the standard way using ESO MIDAS software, as described in ref. \cite{Hainaut2019}. The photometric calibration was obtained from field stars appearing in the Pan-STARRS and SDSS databases, using the colour corrections from the original filters to SDSS and Cron-Cousins systems as described in ref. \cite{tonry12}. The uncertainty reported included the dispersion on the photometric zero-points, the noise from the measurement and the error from sky background subtraction.

The magnitudes of the comet were measured through a series of circular apertures, the residual background being estimated from the median of a large region far from the object. The magnitudes measured using a $5^{\prime\prime}$ radius aperture are reported in Supplementary Table~2. The $R$ magnitudes are converted into $Af\rho$, where $A$ is the dust reflectivity (i.e., albedo), $f$ is the dust filling factor and $\rho$ is the linear radius of the field of view at the comet \cite{ahearn1984AJ}. An order-of-magnitude dust mass loss rate $Q({\rm dust})$ could be derived from $Af\rho$ (see ref.\cite{Cremonese2020}) if we specify the dust velocity $v_d$, the dust mass density $\rho_d$, the mean dust size $\langle a\rangle$ inferred from optical observations, and the dust geometrical albedo $p_V$:
\begin{equation}
   A f\rho \approx \frac{3\ p_V\  Q({\rm dust})}{v_d\,\rho_d\ \langle a\rangle} ~.
\end{equation}
By taking 
$\rho_d=\rhoCD\approx2,650\kg\m^{-3}$, $v_d=3\m\s^{-1}$, 
$\langle a\rangle=1\mm$, and $p_V=0.04$
(ref. \cite{Cremonese2020,Kim2020}),
we derived the dust mass loss rates and listed them in Supplementary Table~2. We note that these are just order-of-magnitude estimates 
since the optical observations typically probe small, micron-sized dust more effectively than mm-sized dust and the mean dust size of $\langle a\rangle=1\mm$ was inferred indirectly from  dynamical models \cite{Cremonese2020,Kim2020}.


\section*{Acknowledgements}
We thank Pei-Ying Hsieh, Chentao Yang for help with the ALMA data, Zahed Wahhaj, David Jewitt, Xuejuan Yang and the referees for their very constructive comments. This paper makes use of the following ALMA data: ADS/JAO.ALMA\#2019.A.00002.S. ALMA is a partnership of ESO (representing its member states), NSF (USA) and NINS (Japan), together with NRC (Canada), MOST and ASIAA (Taiwan), and KASI (Republic of Korea), in cooperation with the Republic of Chile. The Joint ALMA Observatory is operated by ESO, AUI/NRAO and NAOJ. Based on observations collected at the European Organisation for Astronomical Research in the Southern Hemisphere under ESO program 105.205Q.001. A.L. was supported in part by NSF AST-1816411, HST-AR-15037.001-A, and Chandra TM9-20009X. M.A.C. was supported by the National Science Foundation (under Grant No. AST-1614471) and K.J.M. and J.V.K. were supported by NASA (Grant No. 80NSSC18K0853).

\section*{Author contributions statement}
B.Y. led the application and organization of the ALMA observations and led the writing of this paper. A.L. performed dust modeling of the ALMA data and assisted in the writing of this paper and the ALMA proposal. O.R.H. analyzed the FORS data and M.A.C. analyzed the CO data. C.S.C. assisted in writing the ALMA proposal and reduced the ALMA data. J.P.W.  contributed to ALMA observation design and data interpretation. K.J.M., J.V.K. and E.V. were co-Is on the telescope proposals and commented on the manuscript.

\section*{Data availability}
This work makes use of ALMA dataset ADS/JAO.ALMA\#2019.A.00002.S, which is available for download from the ALMA Science Archive (http://almascience.nrao.edu/aq/) following a 9-month proprietary period. The VLT dataset is available for download from the ESO Science Archive (http://archive.eso.org/eso/eso$\_$archive$\_$main.html), under ESO program ID 2103.C-5068 and 0105.C-0250, PI O.R.H., following a 1-year proprietary period.

\section*{Competing interests}
The authors declare no competing financial or non-financial interests.

\begin{thebibliography}{10}
\urlstyle{rm}
\expandafter\ifx\csname url\endcsname\relax
  \def\url#1{\texttt{#1}}\fi
\expandafter\ifx\csname urlprefix\endcsname\relax\def\urlprefix{URL }\fi
\expandafter\ifx\csname doiprefix\endcsname\relax\def\doiprefix{DOI: }\fi
\providecommand{\bibinfo}[2]{#2}
\providecommand{\eprint}[2][]{\url{#2}}

\bibitem{Charnoz2003}
\bibinfo{author}{{Charnoz}, S.} \& \bibinfo{author}{{Morbidelli}, A.}
\newblock \bibinfo{journal}{\bibinfo{title}{{Coupling dynamical and collisional
  evolution of small bodies:. an application to the early ejection of
  planetesimals from the Jupiter-Saturn region}}}.
\newblock {\emph{\JournalTitle{Icarus}}} \textbf{\bibinfo{volume}{166}},
  \bibinfo{pages}{141--156} (\bibinfo{year}{2003}).

\bibitem{meech2017}
\bibinfo{author}{{Meech}, K.~J.} \emph{et~al.}
\newblock \bibinfo{journal}{\bibinfo{title}{{A brief visit from a red and
  extremely elongated interstellar asteroid}}}.
\newblock {\emph{\JournalTitle{Nature}}} \textbf{\bibinfo{volume}{552}},
  \bibinfo{pages}{378--381} (\bibinfo{year}{2017}).

\bibitem{ISSI2019}
\bibinfo{author}{{'Oumuamua ISSI Team}} \emph{et~al.}
\newblock \bibinfo{journal}{\bibinfo{title}{{The natural history of
  `Oumuamua}}}.
\newblock {\emph{\JournalTitle{Nature Astronomy}}}
  \textbf{\bibinfo{volume}{3}}, \bibinfo{pages}{594--602}
  (\bibinfo{year}{2019}).

\bibitem{Guzik2020}
\bibinfo{author}{{Guzik}, P.} \emph{et~al.}
\newblock \bibinfo{journal}{\bibinfo{title}{{Initial characterization of
  interstellar comet 2I/Borisov}}}.
\newblock {\emph{\JournalTitle{Nature Astronomy}}}
  \textbf{\bibinfo{volume}{4}}, \bibinfo{pages}{53--57} (\bibinfo{year}{2020}).

\bibitem{Fitzsimmons2019}
\bibinfo{author}{{Fitzsimmons}, A.} \emph{et~al.}
\newblock \bibinfo{journal}{\bibinfo{title}{{Detection of CN Gas in
  Interstellar Object 2I/Borisov}}}.
\newblock {\emph{\JournalTitle{Astrophys. J. Letter}}}
  \textbf{\bibinfo{volume}{885}}, \bibinfo{pages}{L9} (\bibinfo{year}{2019}).

\bibitem{Opitom2019}
\bibinfo{author}{{Opitom}, C.} \emph{et~al.}
\newblock \bibinfo{journal}{\bibinfo{title}{{2I/Borisov: A C$_{2}$-depleted
  interstellar comet}}}.
\newblock {\emph{\JournalTitle{Astron. Astrophys.}}}
  \textbf{\bibinfo{volume}{631}}, \bibinfo{pages}{L8} (\bibinfo{year}{2019}).

\bibitem{Bodewits2020}
\bibinfo{author}{{Bodewits}, D.} \emph{et~al.}
\newblock \bibinfo{journal}{\bibinfo{title}{{The carbon monoxide-rich
  interstellar comet 2I/Borisov}}}.
\newblock {\emph{\JournalTitle{Nature Astronomy}}}  (\bibinfo{year}{2020}).

\bibitem{Cordiner2020}
\bibinfo{author}{{Cordiner}, M.~A.} \emph{et~al.}
\newblock \bibinfo{journal}{\bibinfo{title}{{Unusually high CO abundance of the
  first active interstellar comet}}}.
\newblock {\emph{\JournalTitle{Nature Astronomy}}}  (\bibinfo{year}{2020}).

\bibitem{Jewitt2020}
\bibinfo{author}{{Jewitt}, D.} \emph{et~al.}
\newblock \bibinfo{journal}{\bibinfo{title}{{The Nucleus of Interstellar Comet
  2I/Borisov}}}.
\newblock {\emph{\JournalTitle{Astrophys. J. Lett.}}}
  \textbf{\bibinfo{volume}{888}}, \bibinfo{pages}{L23} (\bibinfo{year}{2020}).

\bibitem{Hui2020}
\bibinfo{author}{{Hui}, M.-T.}, \bibinfo{author}{{Ye}, Q.-Z.},
  \bibinfo{author}{{F{\"o}hring}, D.}, \bibinfo{author}{{Hung}, D.} \&
  \bibinfo{author}{{Tholen}, D.~J.}
\newblock \bibinfo{journal}{\bibinfo{title}{{Physical Characterization of
  Interstellar Comet 2I/2019 Q4 (Borisov)}}}.
\newblock {\emph{\JournalTitle{Astron. J.}}} \textbf{\bibinfo{volume}{160}},
  \bibinfo{pages}{92} (\bibinfo{year}{2020}).

\bibitem{Cochran2015}
\bibinfo{author}{{Cochran}, A.~L.} \emph{et~al.}
\newblock \bibinfo{journal}{\bibinfo{title}{{The Composition of Comets}}}.
\newblock {\emph{\JournalTitle{Space Science Rev.}}}
  \textbf{\bibinfo{volume}{197}}, \bibinfo{pages}{9--46}
  (\bibinfo{year}{2015}).

\bibitem{Fray2017}
\bibinfo{author}{{Fray}, N.} \emph{et~al.}
\newblock \bibinfo{journal}{\bibinfo{title}{{Nitrogen-to-carbon atomic ratio
  measured by COSIMA in the particles of comet 67P/Churyumov-Gerasimenko}}}.
\newblock {\emph{\JournalTitle{Mon. Not. R. Astron. Soc.}}}
  \textbf{\bibinfo{volume}{469}}, \bibinfo{pages}{S506--S516}
  (\bibinfo{year}{2017}).

\bibitem{Levasseur2018}
\bibinfo{author}{{Levasseur-Regourd}, A.-C.} \emph{et~al.}
\newblock \bibinfo{journal}{\bibinfo{title}{{Cometary Dust}}}.
\newblock {\emph{\JournalTitle{Space Science Rev.}}}
  \textbf{\bibinfo{volume}{214}}, \bibinfo{pages}{64} (\bibinfo{year}{2018}).

\bibitem{Ahearn:2017}
\bibinfo{author}{{A'Hearn}, M.~F.}
\newblock \bibinfo{journal}{\bibinfo{title}{{Comets: looking ahead}}}.
\newblock {\emph{\JournalTitle{Philosophical Transactions of the Royal Society
  of London Series A}}} \textbf{\bibinfo{volume}{375}},
  \bibinfo{pages}{20160261} (\bibinfo{year}{2017}).

\bibitem{Davidsson2016}
\bibinfo{author}{{Davidsson}, B.~J.~R.} \emph{et~al.}
\newblock \bibinfo{journal}{\bibinfo{title}{{The primordial nucleus of comet
  67P/Churyumov-Gerasimenko}}}.
\newblock {\emph{\JournalTitle{Astron. Astrophys.}}}
  \textbf{\bibinfo{volume}{592}}, \bibinfo{pages}{A63} (\bibinfo{year}{2016}).

\bibitem{Fulle2017}
\bibinfo{author}{{Fulle}, M.}
\newblock \bibinfo{journal}{\bibinfo{title}{{The ice content of Kuiper belt
  objects}}}.
\newblock {\emph{\JournalTitle{Nature Astronomy}}}
  \textbf{\bibinfo{volume}{1}}, \bibinfo{pages}{0018} (\bibinfo{year}{2017}).

\bibitem{guttler2019}
\bibinfo{author}{{G{\"u}ttler}, C.} \emph{et~al.}
\newblock \bibinfo{journal}{\bibinfo{title}{{Synthesis of the morphological
  description of cometary dust at comet 67P/Churyumov-Gerasimenko}}}.
\newblock {\emph{\JournalTitle{Astron. Astrophys.}}}
  \textbf{\bibinfo{volume}{630}}, \bibinfo{pages}{A24} (\bibinfo{year}{2019}).

\bibitem{Li1998}
\bibinfo{author}{{Li}, A.} \& \bibinfo{author}{{Greenberg}, J.~M.}
\newblock \bibinfo{journal}{\bibinfo{title}{{From Interstellar Dust to Comets:
  Infrared Emission from Comet Hale-Bopp (C/1995 O1)}}}.
\newblock {\emph{\JournalTitle{Astrophys. J. Letter}}}
  \textbf{\bibinfo{volume}{498}}, \bibinfo{pages}{L83--L87}
  (\bibinfo{year}{1998}).

\bibitem{Wentworth1922}
\bibinfo{author}{{Wentworth}, C.~K.}
\newblock \bibinfo{journal}{\bibinfo{title}{{A Scale of Grade and Class Terms
  for Clastic Sediments}}}.
\newblock {\emph{\JournalTitle{Journal of Geology}}}
  \textbf{\bibinfo{volume}{30}}, \bibinfo{pages}{377--392}
  (\bibinfo{year}{1922}).

\bibitem{Johansen2017}
\bibinfo{author}{{Johansen}, A.} \& \bibinfo{author}{{Lambrechts}, M.}
\newblock \bibinfo{journal}{\bibinfo{title}{{Forming Planets via Pebble
  Accretion}}}.
\newblock {\emph{\JournalTitle{Annual Review of Earth and Planetary Sciences}}}
  \textbf{\bibinfo{volume}{45}}, \bibinfo{pages}{359--387}
  (\bibinfo{year}{2017}).

\bibitem{jewitt1999}
\bibinfo{author}{{Jewitt}, D.} \& \bibinfo{author}{{Matthews}, H.}
\newblock \bibinfo{journal}{\bibinfo{title}{{Particulate Mass Loss from Comet
  Hale-Bopp}}}.
\newblock {\emph{\JournalTitle{Astron. J.}}} \textbf{\bibinfo{volume}{117}},
  \bibinfo{pages}{1056--1062} (\bibinfo{year}{1999}).

\bibitem{Bohren:1983}
\bibinfo{author}{Bohren, C.~F.} \& \bibinfo{author}{Huffman, D.~R.}
\newblock \emph{\bibinfo{title}{Absorption and scattering of light by small
  particles}} (\bibinfo{publisher}{New York: Wiley}, \bibinfo{year}{1983}).

\bibitem{Divine1986}
\bibinfo{author}{{Divine}, N.} \emph{et~al.}
\newblock \bibinfo{journal}{\bibinfo{title}{{The Comet Halley Dust and Gas
  Environment}}}.
\newblock {\emph{\JournalTitle{Space Science Reviews}}}
  \textbf{\bibinfo{volume}{43}}, \bibinfo{pages}{1} (\bibinfo{year}{1986}).

\bibitem{Ye2020}
\bibinfo{author}{{Ye}, Q.} \emph{et~al.}
\newblock \bibinfo{journal}{\bibinfo{title}{{Pre-discovery Activity of New
  Interstellar Comet 2I/Borisov beyond 5 au}}}.
\newblock {\emph{\JournalTitle{Astron. J.}}} \textbf{\bibinfo{volume}{159}},
  \bibinfo{pages}{77} (\bibinfo{year}{2020}).

\bibitem{jewitt2019}
\bibinfo{author}{{Jewitt}, D.} \& \bibinfo{author}{{Luu}, J.}
\newblock \bibinfo{journal}{\bibinfo{title}{{Initial Characterization of
  Interstellar Comet 2I/2019 Q4 (Borisov)}}}.
\newblock {\emph{\JournalTitle{Astrophys. J. Letter}}}
  \textbf{\bibinfo{volume}{886}}, \bibinfo{pages}{L29} (\bibinfo{year}{2019}).

\bibitem{Cremonese2020}
\bibinfo{author}{{Cremonese}, G.} \emph{et~al.}
\newblock \bibinfo{journal}{\bibinfo{title}{{Dust Environment Model of the
  Interstellar Comet 2I/Borisov}}}.
\newblock {\emph{\JournalTitle{Astrophys. J. Letter}}}
  \textbf{\bibinfo{volume}{893}}, \bibinfo{pages}{L12} (\bibinfo{year}{2020}).

\bibitem{Kim2020}
\bibinfo{author}{{Kim}, Y.} \emph{et~al.}
\newblock \bibinfo{journal}{\bibinfo{title}{{Coma Anisotropy and the Rotation
  Pole of Interstellar Comet 2I/Borisov}}}.
\newblock {\emph{\JournalTitle{Astrophys. J. Letter}}}
  \textbf{\bibinfo{volume}{895}}, \bibinfo{pages}{L34} (\bibinfo{year}{2020}).

\bibitem{Patzold2019}
\bibinfo{author}{{P{\"a}tzold}, M.} \emph{et~al.}
\newblock \bibinfo{journal}{\bibinfo{title}{{The Nucleus of comet
  67P/Churyumov-Gerasimenko - Part I: The global view - nucleus mass,
  mass-loss, porosity, and implications}}}.
\newblock {\emph{\JournalTitle{Mon. Not. R. Astron. Soc.}}}
  \textbf{\bibinfo{volume}{483}}, \bibinfo{pages}{2337--2346}
  (\bibinfo{year}{2019}).

\bibitem{Bolin2020a}
\bibinfo{author}{{Bolin}, B.~T.} \emph{et~al.}
\newblock \bibinfo{journal}{\bibinfo{title}{{Characterization of the Nucleus,
  Morphology, and Activity of Interstellar Comet 2I/Borisov by Optical and
  Near-infrared GROWTH, Apache Point, IRTF, ZTF, and Keck Observations}}}.
\newblock {\emph{\JournalTitle{Astron. J}}} \textbf{\bibinfo{volume}{160}},
  \bibinfo{pages}{26} (\bibinfo{year}{2020}).

\bibitem{Kolokolova1997}
\bibinfo{author}{{Kolokolova}, L.}, \bibinfo{author}{{Jockers}, K.},
  \bibinfo{author}{{Chernova}, G.} \& \bibinfo{author}{{Kiselev}, N.}
\newblock \bibinfo{journal}{\bibinfo{title}{{Properties of Cometary Dust from
  Color and Polarization}}}.
\newblock {\emph{\JournalTitle{Icarus}}} \textbf{\bibinfo{volume}{126}},
  \bibinfo{pages}{351--361} (\bibinfo{year}{1997}).

\bibitem{Yang2020}
\bibinfo{author}{{Yang}, B.} \emph{et~al.}
\newblock \bibinfo{journal}{\bibinfo{title}{{Searching for water ice in the
  coma of interstellar object 2I/Borisov}}}.
\newblock {\emph{\JournalTitle{Astron. and Astrophys.}}}
  \textbf{\bibinfo{volume}{634}}, \bibinfo{pages}{L6} (\bibinfo{year}{2020}).

\bibitem{hainaut12MBOSS}
\bibinfo{author}{{Hainaut}, O.~R.}, \bibinfo{author}{{Boehnhardt}, H.} \&
  \bibinfo{author}{{Protopapa}, S.}
\newblock \bibinfo{journal}{\bibinfo{title}{{Colours of minor bodies in the
  outer solar system. II. A statistical analysis revisited}}}.
\newblock {\emph{\JournalTitle{Astron. Astrophys.}}}
  \textbf{\bibinfo{volume}{546}}, \bibinfo{pages}{A115} (\bibinfo{year}{2012}).

\bibitem{xing2020}
\bibinfo{author}{{Xing}, Z.}, \bibinfo{author}{{Bodewits}, D.},
  \bibinfo{author}{{Noonan}, J.} \& \bibinfo{author}{{Bannister}, M.~T.}
\newblock \bibinfo{journal}{\bibinfo{title}{{Water Production Rates and
  Activity of Interstellar Comet 2I/Borisov}}}.
\newblock {\emph{\JournalTitle{Astrophys. J. Letter}}}
  \textbf{\bibinfo{volume}{893}}, \bibinfo{pages}{L48} (\bibinfo{year}{2020}).

\bibitem{lauter2019}
\bibinfo{author}{{L{\"a}uter}, M.}, \bibinfo{author}{{Kramer}, T.},
  \bibinfo{author}{{Rubin}, M.} \& \bibinfo{author}{{Altwegg}, K.}
\newblock \bibinfo{journal}{\bibinfo{title}{{Surface localization of gas
  sources on comet 67P/Churyumov-Gerasimenko based on DFMS/COPS data}}}.
\newblock {\emph{\JournalTitle{Mon. Not. R. Astron. Soc.}}}
  \textbf{\bibinfo{volume}{483}}, \bibinfo{pages}{852--861}
  (\bibinfo{year}{2019}).

\bibitem{Rotundi2015}
\bibinfo{author}{{Rotundi}, A.} \emph{et~al.}
\newblock \bibinfo{journal}{\bibinfo{title}{{Dust measurements in the coma of
  comet 67P/Churyumov-Gerasimenko inbound to the Sun}}}.
\newblock {\emph{\JournalTitle{Science}}} \textbf{\bibinfo{volume}{347}},
  \bibinfo{pages}{aaa3905} (\bibinfo{year}{2015}).

\bibitem{Choukroun2020}
\bibinfo{author}{{Choukroun}, M.} \emph{et~al.}
\newblock \bibinfo{journal}{\bibinfo{title}{{Dust-to-Gas and Refractory-to-Ice
  Mass Ratios of Comet 67P/Churyumov-Gerasimenko from Rosetta Observations}}}.
\newblock {\emph{\JournalTitle{Space Science Reviews}}}
  \textbf{\bibinfo{volume}{216}}, \bibinfo{pages}{44} (\bibinfo{year}{2020}).

\bibitem{Singh1992}
\bibinfo{author}{{Singh}, P.~D.}, \bibinfo{author}{{de Almeida}, A.~A.} \&
  \bibinfo{author}{{Huebner}, W.~F.}
\newblock \bibinfo{journal}{\bibinfo{title}{{Dust Release Rates and Dust-to-Gas
  Mass Ratios of Eight Comets}}}.
\newblock {\emph{\JournalTitle{Astronomical J.}}}
  \textbf{\bibinfo{volume}{104}}, \bibinfo{pages}{848} (\bibinfo{year}{1992}).

\bibitem{Lamy2009}
\bibinfo{author}{{Lamy}, P.~L.}, \bibinfo{author}{{Toth}, I.},
  \bibinfo{author}{{Weaver}, H.~A.}, \bibinfo{author}{{A'Hearn}, M.~F.} \&
  \bibinfo{author}{{Jorda}, L.}
\newblock \bibinfo{journal}{\bibinfo{title}{{Properties of the nuclei and comae
  of 13 ecliptic comets from Hubble Space Telescope snapshot observations}}}.
\newblock {\emph{\JournalTitle{Astronomy and Astrophysics}}}
  \textbf{\bibinfo{volume}{508}}, \bibinfo{pages}{1045--1056}
  (\bibinfo{year}{2009}).

\bibitem{Lorek2016}
\bibinfo{author}{{Lorek}, S.}, \bibinfo{author}{{Gundlach}, B.},
  \bibinfo{author}{{Lacerda}, P.} \& \bibinfo{author}{{Blum}, J.}
\newblock \bibinfo{journal}{\bibinfo{title}{{Comet formation in collapsing
  pebble clouds. What cometary bulk density implies for the cloud mass and
  dust-to-ice ratio}}}.
\newblock {\emph{\JournalTitle{Astron. Astrophys.}}}
  \textbf{\bibinfo{volume}{587}}, \bibinfo{pages}{A128} (\bibinfo{year}{2016}).

\bibitem{Mannel2019}
\bibinfo{author}{{Mannel}, T.} \emph{et~al.}
\newblock \bibinfo{journal}{\bibinfo{title}{{Dust of comet
  67P/Churyumov-Gerasimenko collected by Rosetta/MIDAS: classification and
  extension to the nanometer scale}}}.
\newblock {\emph{\JournalTitle{Astron. Astrophys.}}}
  \textbf{\bibinfo{volume}{630}}, \bibinfo{pages}{A26} (\bibinfo{year}{2019}).

\bibitem{Weidenschilling1977}
\bibinfo{author}{{Weidenschilling}, S.~J.}
\newblock \bibinfo{journal}{\bibinfo{title}{{Aerodynamics of solid bodies in
  the solar nebula.}}}
\newblock {\emph{\JournalTitle{Mon. Not. R. Astron. Soc.}}}
  \textbf{\bibinfo{volume}{180}}, \bibinfo{pages}{57--70}
  (\bibinfo{year}{1977}).

\bibitem{Zsom2010}
\bibinfo{author}{{Zsom}, A.}, \bibinfo{author}{{Ormel}, C.~W.},
  \bibinfo{author}{{G{\"u}ttler}, C.}, \bibinfo{author}{{Blum}, J.} \&
  \bibinfo{author}{{Dullemond}, C.~P.}
\newblock \bibinfo{journal}{\bibinfo{title}{{The outcome of protoplanetary dust
  growth: pebbles, boulders, or planetesimals? II. Introducing the bouncing
  barrier}}}.
\newblock {\emph{\JournalTitle{Astron. Astrophys.}}}
  \textbf{\bibinfo{volume}{513}}, \bibinfo{pages}{A57} (\bibinfo{year}{2010}).

\bibitem{Mukai1983}
\bibinfo{author}{{Mukai}, T.} \& \bibinfo{author}{{Fechtig}, H.}
\newblock \bibinfo{journal}{\bibinfo{title}{{Packing effect of fluffy
  particles}}}.
\newblock {\emph{\JournalTitle{Planetary and Space Science}}}
  \textbf{\bibinfo{volume}{31}}, \bibinfo{pages}{655--658}
  (\bibinfo{year}{1983}).

\bibitem{Feaga2014}
\bibinfo{author}{{Feaga}, L.~M.} \emph{et~al.}
\newblock \bibinfo{journal}{\bibinfo{title}{{Uncorrelated Volatile Behavior
  during the 2011 Apparition of Comet C/2009 P1 Garradd}}}.
\newblock {\emph{\JournalTitle{Astron. J.}}} \textbf{\bibinfo{volume}{147}},
  \bibinfo{pages}{24} (\bibinfo{year}{2014}).

\bibitem{Cooper2003}
\bibinfo{author}{{Cooper}, J.~F.}, \bibinfo{author}{{Christian}, E.~R.},
  \bibinfo{author}{{Richardson}, J.~D.} \& \bibinfo{author}{{Wang}, C.}
\newblock \bibinfo{journal}{\bibinfo{title}{{Proton Irradiation of Centaur,
  Kuiper Belt, and Oort Cloud Objects at Plasma to Cosmic Ray Energy}}}.
\newblock {\emph{\JournalTitle{Earth Moon and Planets}}}
  \textbf{\bibinfo{volume}{92}}, \bibinfo{pages}{261--277}
  (\bibinfo{year}{2003}).

\bibitem{Garrod2019}
\bibinfo{author}{{Garrod}, R.~T.}
\newblock \bibinfo{journal}{\bibinfo{title}{{Simulations of Ice Chemistry in
  Cometary Nuclei}}}.
\newblock {\emph{\JournalTitle{Astrophys. J.}}} \textbf{\bibinfo{volume}{884}},
  \bibinfo{pages}{69} (\bibinfo{year}{2019}).

\bibitem{Jehin2020}
\bibinfo{author}{{Jehin}, E.} \emph{et~al.}
\newblock \bibinfo{title}{{Monitoring of the optical spectrum of comet
  2I/Borisov at the VLT}}.
\newblock In \emph{\bibinfo{booktitle}{European Planetary Science Congress}},
  \bibinfo{pages}{EPSC2020--653} (\bibinfo{year}{2020}).

\bibitem{Bolin2020b}
\bibinfo{author}{{Bolin}, B.~T.} \& \bibinfo{author}{{Lisse}, C.~M.}
\newblock \bibinfo{journal}{\bibinfo{title}{{Constraints on the spin-pole
  orientation, jet morphology, and rotation of interstellar comet 2I/Borisov
  with deep HST imaging}}}.
\newblock {\emph{\JournalTitle{Mon. Not. R. Astron. Soc.}}}
  \textbf{\bibinfo{volume}{497}}, \bibinfo{pages}{4031--4041}
  (\bibinfo{year}{2020}).

\bibitem{walsh2011}
\bibinfo{author}{{Walsh}, K.~J.}, \bibinfo{author}{{Morbidelli}, A.},
  \bibinfo{author}{{Raymond}, S.~N.}, \bibinfo{author}{{O'Brien}, D.~P.} \&
  \bibinfo{author}{{Mandell}, A.~M.}
\newblock \bibinfo{journal}{\bibinfo{title}{{A low mass for Mars from Jupiter's
  early gas-driven migration}}}.
\newblock {\emph{\JournalTitle{Nature}}} \textbf{\bibinfo{volume}{475}},
  \bibinfo{pages}{206--209} (\bibinfo{year}{2011}).

\bibitem{Batalha2014}
\bibinfo{author}{{Batalha}, N.~M.}
\newblock \bibinfo{journal}{\bibinfo{title}{{Exploring exoplanet populations
  with NASA's Kepler Mission}}}.
\newblock {\emph{\JournalTitle{Proceedings of the National Academy of
  Science}}} \textbf{\bibinfo{volume}{111}}, \bibinfo{pages}{12647--12654}
  (\bibinfo{year}{2014}).

\bibitem{Kolokolova2004}
\bibinfo{author}{{Kolokolova}, L.}, \bibinfo{author}{{Hanner}, M.~S.},
  \bibinfo{author}{{Levasseur-Regourd}, A.~C.} \& \bibinfo{author}{{Gustafson},
  B. {\r{A}}.~S.}
\newblock \emph{\bibinfo{title}{{Physical properties of cometary dust from
  light scattering and thermal emission}}}, \bibinfo{pages}{Comets II, 577}
  (\bibinfo{publisher}{University of Arizona Press, Tucson},
  \bibinfo{year}{2004}).

\bibitem{Kimura2016}
\bibinfo{author}{{Kimura}, H.}, \bibinfo{author}{{Kolokolova}, L.},
  \bibinfo{author}{{Li}, A.} \& \bibinfo{author}{{Lebreton}, J.}
\newblock \emph{\bibinfo{title}{{Light Scattering and Thermal Emission by
  Primitive Dust Particles in Planetary Systems}}}, \bibinfo{pages}{Light
  Scattering Reviews, 363--418} (\bibinfo{publisher}{Springer},
  \bibinfo{year}{2016}).

\bibitem{Draine:1984zt}
\bibinfo{author}{Draine, B.~T.} \& \bibinfo{author}{Lee, H.~M.}
\newblock \bibinfo{journal}{\bibinfo{title}{Optical properties of interstellar
  graphite and silicate grains}}.
\newblock {\emph{\JournalTitle{Astrophysical Journal}}}
  \textbf{\bibinfo{volume}{285}}, \bibinfo{pages}{89--108}
  (\bibinfo{year}{1984}).

\bibitem{Li1997}
\bibinfo{author}{{Li}, A.} \& \bibinfo{author}{{Greenberg}, J.~M.}
\newblock \bibinfo{journal}{\bibinfo{title}{{A unified model of interstellar
  dust.}}}
\newblock {\emph{\JournalTitle{Astron. Astrophys.}}}
  \textbf{\bibinfo{volume}{323}}, \bibinfo{pages}{566--584}
  (\bibinfo{year}{1997}).

\bibitem{Bardyn2017}
\bibinfo{author}{{Bardyn}, A.} \emph{et~al.}
\newblock \bibinfo{journal}{\bibinfo{title}{{Carbon-rich dust in comet
  67P/Churyumov-Gerasimenko measured by COSIMA/Rosetta}}}.
\newblock {\emph{\JournalTitle{Mon. Not. R. Astron. Soc.}}}
  \textbf{\bibinfo{volume}{469}}, \bibinfo{pages}{S712--S722}
  (\bibinfo{year}{2017}).

\bibitem{Jessberger1988}
\bibinfo{author}{{Jessberger}, E.~K.}, \bibinfo{author}{{Christoforidis}, A.}
  \& \bibinfo{author}{{Kissel}, J.}
\newblock \bibinfo{journal}{\bibinfo{title}{{Aspects of the major element
  composition of Halley's dust}}}.
\newblock {\emph{\JournalTitle{Nature}}} \textbf{\bibinfo{volume}{332}},
  \bibinfo{pages}{691--695} (\bibinfo{year}{1988}).

\bibitem{Li2003}
\bibinfo{author}{{Li}, A.} \& \bibinfo{author}{{Lunine}, J.~I.}
\newblock \bibinfo{journal}{\bibinfo{title}{{Modeling the Infrared Emission
  from the HD 141569A Disk}}}.
\newblock {\emph{\JournalTitle{Astrophys. J.}}} \textbf{\bibinfo{volume}{594}},
  \bibinfo{pages}{987--1010} (\bibinfo{year}{2003}).

\bibitem{Lasue2009}
\bibinfo{author}{{Lasue}, J.}, \bibinfo{author}{{Levasseur-Regourd}, A.~C.},
  \bibinfo{author}{{Hadamcik}, E.} \& \bibinfo{author}{{Alcouffe}, G.}
\newblock \bibinfo{journal}{\bibinfo{title}{{Cometary dust properties retrieved
  from polarization observations: Application to C/1995 O1 Hale Bopp and
  1P/Halley}}}.
\newblock {\emph{\JournalTitle{Icarus}}} \textbf{\bibinfo{volume}{199}},
  \bibinfo{pages}{129--144} (\bibinfo{year}{2009}).

\bibitem{Draine1994}
\bibinfo{author}{{Draine}, B.~T.} \& \bibinfo{author}{{Flatau}, P.~J.}
\newblock \bibinfo{journal}{\bibinfo{title}{{Discrete-dipole approximation for
  scattering calculations}}}.
\newblock {\emph{\JournalTitle{Journal of the Optical Society of America A}}}
  \textbf{\bibinfo{volume}{11}}, \bibinfo{pages}{1491--1499}
  (\bibinfo{year}{1994}).

\bibitem{Hainaut2019}
\bibinfo{author}{{Hainaut}, O.~R.} \emph{et~al.}
\newblock \bibinfo{journal}{\bibinfo{title}{{Disintegration of active asteroid
  P/2016 G1 (PANSTARRS)}}}.
\newblock {\emph{\JournalTitle{Astron. Astrophys.}}}
  \textbf{\bibinfo{volume}{628}}, \bibinfo{pages}{A48} (\bibinfo{year}{2019}).

\bibitem{tonry12}
\bibinfo{author}{{Tonry}, J.~L.} \emph{et~al.}
\newblock \bibinfo{journal}{\bibinfo{title}{{The Pan-STARRS1 Photometric
  System}}}.
\newblock {\emph{\JournalTitle{Astrophys. J.}}} \textbf{\bibinfo{volume}{750}},
  \bibinfo{pages}{99} (\bibinfo{year}{2012}).

\bibitem{ahearn1984AJ}
\bibinfo{author}{{A'Hearn}, M.~F.}, \bibinfo{author}{{Schleicher}, D.~G.},
  \bibinfo{author}{{Millis}, R.~L.}, \bibinfo{author}{{Feldman}, P.~D.} \&
  \bibinfo{author}{{Thompson}, D.~T.}
\newblock \bibinfo{journal}{\bibinfo{title}{{Comet Bowell 1980b}}}.
\newblock {\emph{\JournalTitle{Astron. J.}}} \textbf{\bibinfo{volume}{89}},
  \bibinfo{pages}{579--591} (\bibinfo{year}{1984}).

\end{thebibliography}

\section*{Supplementary information}
 -----------------	
 
 \newcommand{\beginsupplement}{%
        \setcounter{table}{0}
        \renewcommand{\thetable}{S\arabic{table}}%
        \setcounter{figure}{0}
        \renewcommand{\thefigure}{S\arabic{figure}}%
     }
\beginsupplement
\begin{table*}
    \centering
\begin{tabular}{llrrrrrrrrrr}
\hline
Date       & Filt.&$N$& $t$ & $\rh$ & $\Delta$ & G. lat. & TA & PlAng & STO & PsAng& PsAMV \\
UT         &      &  &[s]&[au] & [au]     & [deg]   &[deg]&[deg] &[deg]& [deg]& [deg] \\
\hline
\hline
2019-Nov-24 	  &$R$  &3   &21   &2.0318 &2.1126 &46.7 &349.7 &-17.4 &27.5 &289.1 &330.3 \\
		  &$b$	&3   &60\\
		  &$v$	&3   &30\\
		  &$I$  &3   &24\\
		  &$z$	&3   &45\\
2019-Dec-01 	  &$R$  &7   &50   &2.0132 &2.0467 &43.9 &354.7 &-16.8 &28.1 &289.2 &329.4 \\
2019-Dec-03 	  &$R$  &7   &50   &2.0101 &2.0307 &43.0 &356.1 &-16.6 &28.2 &289.3 &329.0 \\
2019-Dec-26 	  &$R$  &3   &21   &2.0445 &1.9372 &29.6 &12.6 &-11.5 &28.4 &292.5 &322.7 \\
		  &$b$	&3   &60\\
		  &$v$	&3   &30\\
		  &$I$  &3   &24\\
		  &$z$	&3   &45\\		  
2019-Dec-27 	  &$R$  &21  &147  &2.0489 &1.9369 &29.0 &13.3 &-11.2 &28.4 &292.7 &322.4 \\
2020-Jan-27	      &$R$	&28	 &196	&2.2895	&2.0441	&10.1&32.9	&-0.55	&25.4	&303.9	&305.7	\\
2020-Feb-18 	  &$R$  &28  &2520 &2.5560 &2.2149 &0.6 &43.9 &6.3 &22.5 &317.3 &292.8 \\
2020-Feb-19 	  &$R$  &7   &490  &2.5696 &2.2238 &0.2 &44.3 &6.5 &22.3 &318.1 &292.4 \\
		  &$b$	&3   &600\\
		  &$I$  &3   &240\\
		  &$v$	&3   &300\\
		  &$z$	&3   &450\\		  
2020-Feb-27 	  &$R$  &55  &4850 &2.6820 &2.2983 &-2.2 &47.7 &8.5 &21.2 &324.5 &289.0 \\
2020-Feb-29 	  &$R$  &112 &10080 &2.7111 &2.3177 &-2.8 &48.5 &9.0 &20.9 &326.2 &288.4 \\
2020-Mar-01 	  &$R$  &28  &2520 &2.7257 &2.3276 &-3.0 &48.9 &9.2 &20.8 &327.1 &288.0 \\
\hline
\end{tabular}
\caption{Circumstances and geometry of the VLT FORS2 observations. Notes: $N$ is the number of exposures; $t$ is the total exposure time; $\rh$ and $\Delta$ are respectively the helio- and geocentric distances; G. lat. is the galactic latitude, TA is the true anomaly; PlAng is the angle between the observer and the orbital plane; STO is the Sun-Target-Observer angle, i.e. the solar phase; PsAng is the position angle of the extended Sun-Target vector; PsAMV is the position angle of the negative of the target's heliocentric velocity vector.}
    \label{tab:obsvlt}
\end{table*}
\begin{table*}
    \centering

\begin{tabular}{lcccccccc}
\hline
Epoch      & $b$       & $v$  & $R$      & $I$             & $z$             &$S$ & $Af\rho$ & $Q({\rm dust})$\\
(UT)       &     &    &     &     &   &(\%/100$\nm$) & (m) & ($\kg\s^{-1}$) \\
\hline
\hline
2019-Nov-24& 17.97 $\pm$ 0.03& 17.19 $\pm$ 0.02& 16.72 $\pm$ 0.02& 16.29 $\pm$ 0.03& 16.53 $\pm$ 0.03& 11.0 $\pm$ 0.9 & 0.63 $\pm$ 0.03 & 42 \\
2019-Dec-01&   &                     	       & 16.69 $\pm$ 0.03&                 &                 &                & 0.62 $\pm$ 0.03 & 41 \\
2019-Dec-03&   &                    	       & 16.67 $\pm$ 0.03&                 &                 &                & 0.63 $\pm$ 0.03 & 42 \\
2019-Dec-26& 18.07 $\pm$ 0.03& 17.29 $\pm$ 0.02& 16.82 $\pm$ 0.03& 16.36 $\pm$ 0.03& 16.63 $\pm$ 0.03& 11.4 $\pm$ 1.0 & 0.54 $\pm$ 0.03 & 36 \\
2019-Dec-27&   &                    	       & 16.82 $\pm$ 0.03&                 &                 &                & 0.54 $\pm$ 0.03 & 35 \\
2020-Jan-27&   &                    	       & 17.22 $\pm$ 0.05&                 &                 &                & 0.49 $\pm$ 0.03 & 32 \\
2020-Feb-18&   &                    	       & 17.48 $\pm$ 0.05&                 &                 &                & 0.52 $\pm$ 0.03 & 34 \\
2020-Feb-19& 18.96 $\pm$ 0.03& 17.95 $\pm$ 0.02& 17.52 $\pm$ 0.03& 16.58 $\pm$ 0.03& 17.40 $\pm$ 0.05&  9.3 $\pm$ 3.0 & 0.52 $\pm$ 0.02 & 34 \\
2020-Feb-27&   &                    	       & 17.51 $\pm$ 0.05&                 &                 &                & 0.58 $\pm$ 0.03 & 38 \\
2020-Feb-29&   &                    	       & 17.62 $\pm$ 0.05&                 &                 &                & 0.54 $\pm$ 0.04 & 36 \\
2020-Mar-01&   &                    	       & 17.55 $\pm$ 0.06&                 &                 &                & 0.59 $\pm$ 0.04 & 40 \\
\hline
\end{tabular}

\caption{Photometry of 2I/Borisov through an aperture of 5$^{\prime\prime}$ radius. $S^\prime$ is the slope of the spectral reflectivity gradient (normalized to 550$\nm$). }
\label{tab:colors}
\end{table*}


\end{document}